# Reversible Transient Nucleation in Ionic Solutions as the Precursor of Ion Crystallization

Gan Ren (任淦) and Yanting Wang (王延颋) *

*State Key Laboratory of Theoretical Physics, Institute of Theoretical Physics and Kavli Institute for Theoretical Physics China, Chinese Academy of Sciences, 55 East Zhongguancun Road, P. O. Box 2735 Beijing, 100190 China*

**ABSTRACT** Molecular dynamics simulations for aqueous sodium chloride solutions were carried out at various concentrations. Supplementary to the Debye-Hückel theory, reversible transient nucleation of ions was observed even in dilute solutions. The average size of formed ion clusters and the lifetime of ion pairs increases with concentration until the saturation point, when ion clusters become stable and individual ions adjust their positions to form ordered lattice structures, leading to irreversible ion crystallization, which is beyond the description of the classical nucleation theory.

Ionic solutions play an essential role in many areas, such as biology and chemistry.[1-3] Although thermodynamic and transport properties of ionic solutions, especially dilute solutions, have been extensively investigated,[4] majority of the studies have focused on their macroscopic properties in an ensemble-averaged fashion. Theoretically, the Debye-Hückel theory[5] well describes ensemble-averaged properties of the extremely dilute solutions and its extended version can be applied to concentrated ionic solutions[6-8] as well as the hard-sphere ionic fluids.[9,10] However, its microscopic structures, nucleation mechanism, crystallization process, and fast dynamics are still not thoroughly understood. In particular, the instantaneous fluctuations of ionic solutions around the statistical ensemble-averaged values, which are important for a better understanding of many biological and chemical processes, are not yet adequately investigated.

In recent years, several experiments have directly or indirectly observed ion cluster formation in ionic solutions. A Raman spectroscopy study of sodium nitrate solutions revealed ion clustering in unsaturated, saturated, and supersaturated solutions, suggesting that the coalescence of clusters results in ion nucleation.[11] A combined ultrafast 2D IR and pump/probe study of vibrational energy transfers discovered the cluster formation in thiocyanate metal solutions.[12] Calcium carbonate cluster formation was also observed by transmission electron microscopy.[13] In addition, molecular dynamics (MD) simulations discovered the cluster formation in sodium chloride solutions under various conditions,[14-16] and their physical properties and nucleation mechanisms were studied.[17-19] MD simulations also suggested that stable ion clusters in sodium chloride solutions may serve as the nuclei for ion nucleation.[20] In calcium carbonate solutions, the stable amorphous calcium carbonate,[21] liquid-like ionic nanoscale polymer structure,[22] and liquid-liquid separation[23] have been observed, suggesting a completely non-classical nucleation for this system. Despite those studies on ion clusters and their relation with ion nucleation, no microscopic mechanisms have yet been provided to systematically explain their molecular origin and their relation with the microscopic fluctuations in ionic solutions remains elusive.

Meanwhile, details of the ion crystallization process in ionic solutions, a special case of nucleation processes, have not been fully understood. The best known and widely used classical nucleation theory (CNT)[24,25] can provide qualitative descriptions for nucleation processes. However, it suffers from a faster nucleation time than experiment for most systems,[26] because in the CNT, clusters are treated as spherical droplets without interior structures and the growth of clusters is simplified to be a process with a single ion attached to or departed from a stationary nucleus at each time. More sophisticated theories, such as the Dillmann-Meier theory,[27] are still not good at describing complex systems like ionic solutions.

In this Letter, we report our work on MD simulations of aqueous sodium chloride solutions at a wide range of concentrations from very dilute 0.09 M to 6.24 M well above the saturation concentration. The purpose of our current study is investigating: (1) instantaneous fluctuations in ionic solutions beyond the description of the Debye-Hückel theory and its extensions; (2) ion clustering in various concentrations; (3) ion crystallization process above saturation. Our simulation results suggest that, below saturation, ions instantaneously aggregate to form ion clusters in different places, but the clusters quickly segregate due to thermal energy and then ions reform new clusters in other places. This reversible transient nucleation of ions happens even in very dilute solutions, but the ensemble average over all instantaneous configurations is always uniform below saturation, to which the Debye-Hückel type theories apply. With increasing concentration, the average cluster size and lifetime of ion pairs increase until above saturation, when the ion clusters are large enough to overcome the thermal energy and become stable. The stable clusters serve as nuclei with more ions continuously joining them and the irreversible ion crystallization happens, when the ions in the amorphous clusters adjust their positions to form ordered lattice structures. This observed ion crystallization is likely a multistep nucleation process beyond the description of the CNT and other existing nucleation theories. It might share theoretical fundamentals with other multistep nucleation processes, such as crystallization process in organic molecular systems,[13, 23, 28] dendritic structure formation of silver nanoclusters,[29] and self-assembly of short peptides.[30]

Aqueous sodium chloride solutions at different concentrations $c$ = 0.09, 0.92, 2.14, 3.60, 4.41, 5.45, and 6.24 M were simulated by using the Gromacs 4.5.4 software package [31, 32] with the periodic boundary condition (PBC) applied. The particle-mesh Ewald algorithm[33] was employed to calculate the long-range electrostatic interactions. The CHARMM force field[34] and the TIP3P water model[35] were used to model the ionic solutions. The time step was 1 fs and the Nosé-Hoover thermostat[36,37] was employed to keep the system temperature constant at $T$ = 300 K. The number of water molecules is 3000 and the concentration is varied by including different numbers of ion pairs. After equilibration, $3 \times 10^4$ different configurations were sampled for each concentration from a 60-ns $NVT$ MD simulation. All the simulations were also performed with the AMBER[38] and OPLS-AA[39] force fields and they all reached the same major qualitative conclusions.

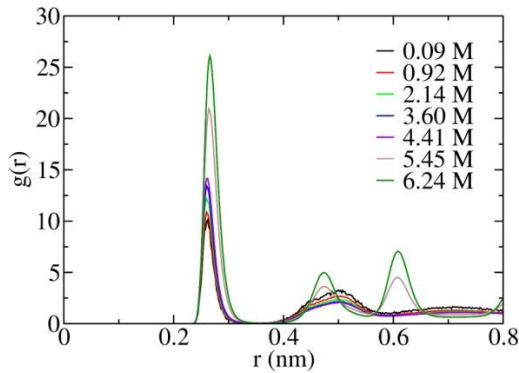

**FIG. 1.** $Na^+$-$Cl^-$ radial distribution functions at various concentrations.

The calculated radial distribution functions (RDFs) between $Na^+$ and $Cl^-$ at various concentrations are plotted in Fig. 1. The shapes of the RDFs do not differ much from 0.09 M to 4.41 M, but change a lot from 4.41 M to 5.45 M, showing that the saturation concentration of the aqueous NaCl solution given by the CHARMM force field is around 5.45 M, close to the experimental value of 5.3 M.[40] The high first peak of the RDF at 0.09 M is attributed to the existence of ion clusters even in dilute solutions, indicating that the Debye-Hückel type theories can only describe ionic solutions in the sense of ensemble average without instantaneous microscopic fluctuations. The crystalline features in the RDF at $c$ = 5.45 M demonstrate that ions form ordered structures above the saturation concentration.

The residence-time correlation functions at various concentrations characterizing the lifetime of ion pairs are shown in Fig. 2. The residence-time correlation function is defined as $C(t) = \langle p(0) p(t) \rangle$, where $p(t)$ is 1 if a given ion is still in the same ion shell at time $t$ and 0 otherwise, and the angular bracket represents the ensemble average over all sampled trajectories. With increasing concentration, the lifetime of ion pairs increases, demonstrating better stability of ion clusters due to their larger sizes. The lifetime drastically increases from $c$ = 4.41 to 5.45 M, indicating a transition from below saturation to above saturation, which is consistent with the RDFs. At the very high concentration of 6.24 M, the lifetime of ion pairs becomes extremely long since most ions form irreversible stable structures.

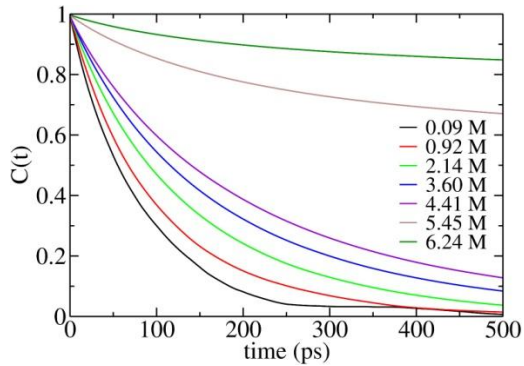

**FIG. 2.** Residence-time correlation functions for the NaCl solutions at various concentrations.

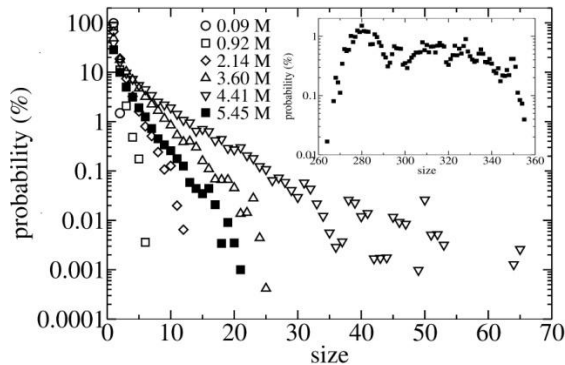

**FIG. 3.** Cluster size distributions for aqueous NaCl solutions at various concentrations. The inset is the size distribution of large clusters at 5.45 M.

The ion clusters were then identified to see how their sizes change with concentration. A cluster is defined as a set of ions with each ion connected with at least one other ion in the same cluster regardless of its charge and species. Two ions are considered connected if they are separated by a distance smaller than a certain cutoff. In this paper, 3.6 Å, the first minimum of the Na-Cl RDF, was used as the cutoff for connected ions. The probability of the appearance of a certain cluster size $i$ is defined as

$p_i = \frac{n_i}{N} \times 100\%$, where $n_i$ is the number of ions composed of clusters with size $i$ and $N$ is the total number of ions. The cluster size distributions at different concentrations are shown in Fig. 3. Small ion clusters form even in solutions as dilute as 0.09 M but with a small probability of about 1%. With increasing concentration, the number of single ions decreases, indicating that more ions incline to form ion clusters. Moreover, as the concentration increases, the probability of the appearance of larger clusters increases. When the concentration is above saturation, a majority of ions form large ion clusters.

The lattice heterogeneity order parameter (LHOP), which had been successfully applied to ionic liquids,[41] was employed to characterize the structural fluctuations in ionic solutions. The heterogeneity order parameter (HOP) is defined as $h = \frac{1}{N} \sum_{i=1}^{N} \sum_{j=1}^{N} \exp\left(-r_{ij}^2 / 2\sigma^2\right)$, where $r_{ij}$ is the distance between ion $i$ and ion $j$ corrected by the PBC, and $\sigma = L/N^{1/3}$ with $L$ the side length of the cubic simulation box and $N$ the total number of ions in each cell. The LHOP, whose detailed definition can be found in ref 42, quantifies the spatial heterogeneity in a lattice space by averaging the HOPs of all particles in the lattice. In our calculations, the simulation space was divided into 6×6×6 lattices and the LHOP value was calculated for each lattice. The LHOPs for two randomly chosen instantaneous configurations are visualized in the first two columns in Fig. 4, and the third column illustrates the LHOPs for the ensemble-averaged configurations over 1000 sampled configurations. The three rows in Fig. 4 represent the results for the concentrations of 0.92, 4.41, and 5.45 M, respectively. Each cell is represented by a small sphere located in the center and colored according to the LHOP value of the cell. The color scale ranges from red to blue with colder colors representing larger LHOP values.

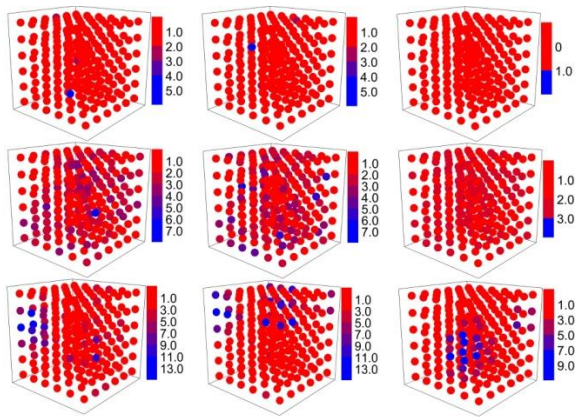

**FIG. 4.** Instantaneous LHOP distributions (left two columns) and their ensemble averages (third column) for the NaCl solution with a concentration of 0.92 (first row), 4.41 (second row), or 5.45 M (third row).

The instantaneous LHOPs for all three concentrations show that the ions are non-uniformly distributed in the simulation box, and the distribution is more non-uniform for a higher concentration. The instantaneous LHOP is different from time to time, demonstrating that ions transiently associate and dissociate driven by thermal fluctuations, but the ensemble-averaged LHOPs are much smaller than the instantaneous LHOPs. For the unsaturated solutions $c = 0.92$ and 4.41 M, the ensemble-averaged LHOPs are both close to zero for all cells, indicating that the ensemble-averaged ion positions distribute uniformly and no stable nuclei exist in the solution. In contrast, for the saturated solution $c = 5.45$ M, the ensemble-averaged LHOP is still non-uniformly distributed, indicating that stable ion clusters exist and serve as the nuclei for ion crystallization. Similar instantaneous microscopic structural fluctuations have also been observed in gold nanoclusters[42] and ionic liquids.[41]

As we have discussed, the CNT is too simple to explain the nucleation process of real systems except several extremely simple systems,[26, 27] while some experiments and simulations[13, 20, 43, 44] have shown that, during nucleation, macromolecules or ions are likely to form intermediate structures before they form ordered structures. In order to study the nucleation process for ion crystallization, we performed non-equilibrium MD simulations for the NaCl solution at $c = 5.45$ M for 125 ns, starting from an initial configuration with the ions randomly distributed in the cubic simulation box. Four consequently sampled snapshots are shown in Fig. 5. At the beginning, the ions are randomly distributed in space. As time goes on, ions first associate to form disordered clusters, and then the larger clusters continue to grow and the smaller clusters shrink. When some clusters become large enough, they are thermodynamically stable and automatically reorganize to form ordered structures. The ordered ion clusters serve as the nuclei and more ions irreversibly join the nuclei to crystallize. This observation agrees with Hassan's detailed simulation study of the ion aggregation and fusion in a NaCl solution,[19] and seems to fall into the description of the two-step nucleation model proposed by Erdemir et al.,[45] in which monomers first associate to form large enough clusters and then the clusters self-organize to form ordered structures.

In summary, by using MD simulation, we have discovered the importance of transient ion nucleation as the microscopic fluctuations in ionic solutions. The reversible transient nucleation was observed even at very dilute concentrations. With increasing concentration, the transiently formed ion clusters have larger sizes and longer lifetimes. Nevertheless, the ensemble average of instantaneous spatial configurations remains uniform below the saturation concentration. Above the saturation concentration, adequately large ion clusters appear which overcomes the thermal energy perturbation and serve as stable nuclei for irreversible nucleation. The ion crystallization is a multistep hierarchical process combining this nucleation process with the self-adjustment of ion positions. This work reveals that the microscopic fluctuations in ionic solutions, which are not described in the Debye-Hückel type theories, are the molecular origin for ion crystallization. The most recent simulation study by the Kathmann group[46] investigating the charge and electric field fluctuations in concentrated aqueous NaCl solutions seems to be supportive to our conclusions. Although this work has only studied aqueous sodium chloride solutions, the discovered mechanisms and qualitative theoretical framework should be independent of specific force fields and can be applied to other regular ionic solutions. The microscopic fluctuations in ionic solutions may help us better understand various chemical and biological processes involving ionic solutions.

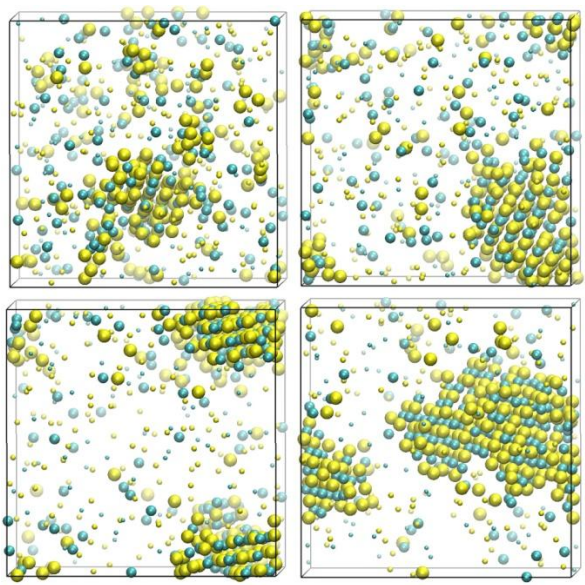

**FIG. 5.** Four consecutive snapshots of an aqueous NaCl solution with a concentration of 5.45 M during a non-equilibrium nucleation process. Larger spheres represent those ions forming clusters.

Funding by the National Basic Research Program of China (973 program, No. 2013CB932804), the National Natural Science Foundation of China (Nos. 91227115 and 11121403) and the Hundred Talent Program of the Chinese Academy of Sciences (CAS), and allocations of computer time from the Supercomputing Center in the Computer Network Information Center at the CAS are gratefully acknowledged.

―――

* Corresponding author: wangyt@itp.ac.cn.